\documentclass{iaus}
\usepackage{graphicx}

\title[Extrasolar Habitable Planet Formation] 
{Habitable Planet Formation in Extreme Planetary Systems: Systems
with Multiple Stars and/or Multiple Planets}

\author[N. Haghighipour]   
{Nader Haghighipour}

\affiliation{Institute for Astronomy and NASA Astrobiology Institute,\\
University of Hawaii-Manoa, Honolulu, HI}

\pubyear{2008}
\volume{IAU249}  
\pagerange{}
\jname{Exoplanets: Detection, Formation and Dynamics}
\editors{}
\begin{document}

\maketitle

\begin{abstract}

Understanding the formation and dynamical evolution of habitable
planets in extrasolar planetary systems is a challenging task.
In this respect, systems with multiple giant 
planets and/or multiple stars present special complications.
The formation of habitable planets in these environments
is strongly affected by the dynamics of their giant planets 
and/or their stellar companions. These objects have
profound effects on the structure of the disk of planetesimals and
protoplanetary objects in which terrestrial-class planets are formed. 
To what extent the current theories of planet formation can be applied
to such {\it "extreme"} planetary systems depends on the dynamical 
characteristics of their planets and/or their binary stars.
In this paper, I present the results of a study of the possibility of the 
existence of Earth-like objects in 
systems with multiple giant planets (namely $\upsilon$ Andromedae,
47 UMa, GJ 876, and 55 Cnc) and discuss the dynamics of the newly 
discovered Neptune-size object in 55 Cnc system.
I will also review habitable planet formation in binary systems
and present the results of a systematic search of the parameter-space
for which Earth-like objects can form and maintain long-term stable orbits
in the habitable zones of binary stars.


\end{abstract}

\firstsection 

\section{Introduction}

The discovery of extrasolar planets during the past decade has confronted 
astronomers with many new challenges. The diverse and surprising dynamical 
characteristics of many of these objects have made scientists wonder to 
what extent the current theories of planet formation can be applied to 
other planetary systems. A major challenge of planetary science is now to 
explain how such planets were formed, how they acquired their unfamiliar 
dynamical state, and whether they can be habitable.

Among the unfamiliar characteristics of the currently known extrasolar
planetary systems, the existence of systems with multiple planets 
in which Jovian-type bodies are in eccentric and 
close-in orbits, and the existence of Jupiter-like planets in 
multi-star systems are particularly interesting.
As shown in figure 1, at the present, 26 extrasolar planetary systems 
contain more than one giant planet. Also, as shown in Table 1, more than 
20\% of planet-hosting stars are members of binary systems. The formation
of terrestrial-class objects in such planetary systems, and the possibility
of their long-term stability in the habitable zones of their host stars
are strongly affected by the dynamical perturbations of the giant planets
and/or the stellar companions. Whether such {\it ``extreme''}
planetary environments can be potential hosts to habitable planets
is the subject of this paper. I will review the possibility of the long-term
stability of terrestrial-class objects in some of multi-planet systems,
and review the current status of research on planet formation in
dual-star environments.

\begin{figure}
\begin{center}
\includegraphics[width=5in]{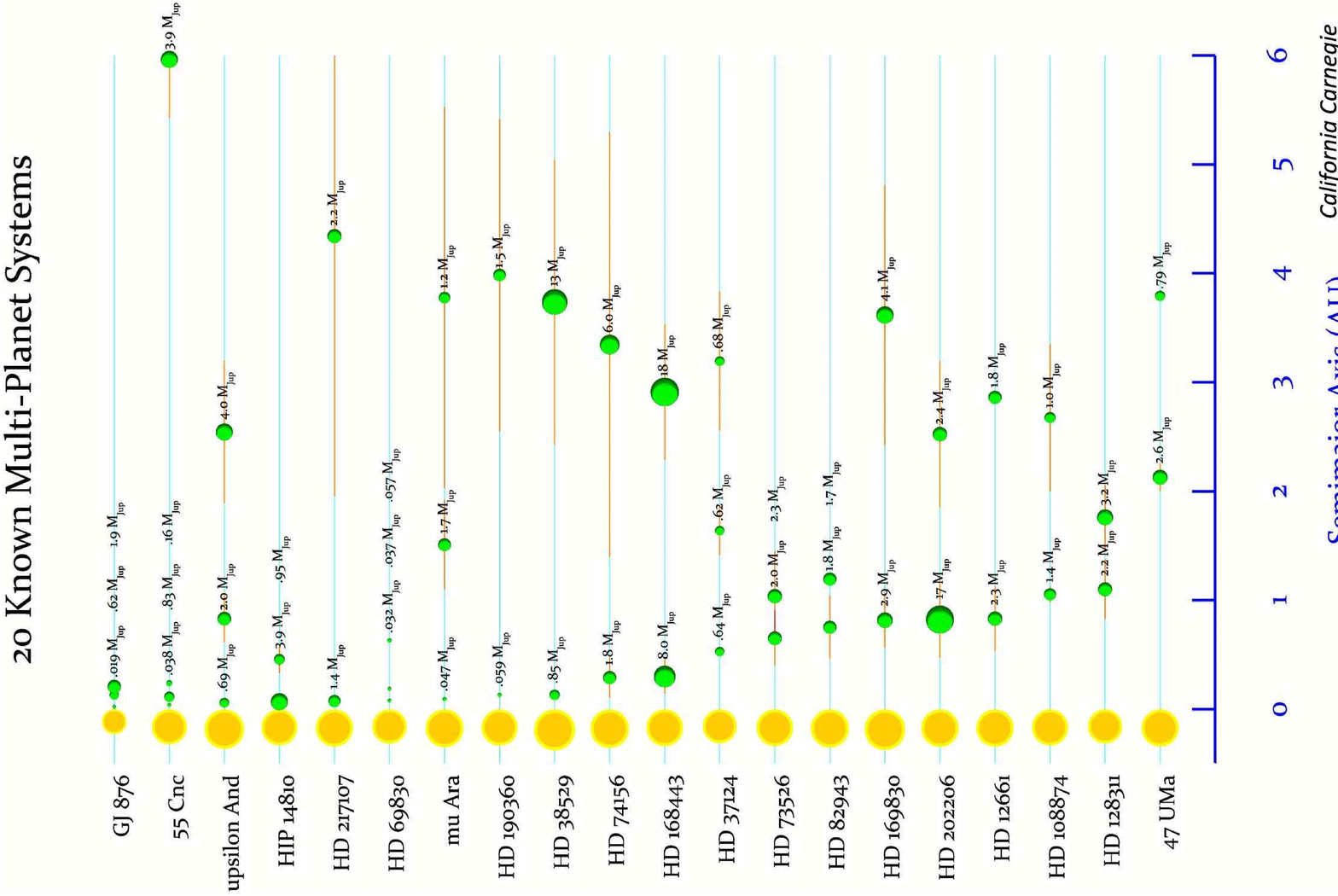} 
\end{center}
\vspace*{-0.5 cm}
\caption{Currently known multi-planet extrasolar planetary systems
(California-Carnegie Planet Search).
\label{fig1}}
\end{figure}

\section{Habitability of Extrasolar Multi-planet Systems}

In order for a planetary system to be habitable, an
Earth-like planet has to maintain its 
orbit in the habitable zone (HZ) of the system's central star for a long time.
This condition requires that the orbital eccentricity of a habitable planet to
be close to zero and its interactions with other bodies of the system 
do not disturb 
its long-term stability. In a multi-planet system, these conditions may not be
easily satisfied. The dynamics of an object in such systems 
is strongly affected by other planets and the habitability of a terrestrial
planet may be influenced by the perturbations from giant bodies. 
The latter is more
significant in systems where the orbits of giant planets are close to the 
habitable zone. The planetary systems of 
$\upsilon$ Andromedae (HZ=1.68-2 AU), 47 UMa (HZ=1.16-1.41 AU), 
GJ 876 (HZ=0.1-0.13 AU), and 55 Cancri (HZ=0.72-0.87 AU) 
are of this kind. In a recent article 
\cite[(Rivera \& Haghighipour, 2007)]{Rivera07}, we studied the stability of
terrestrial-class objects in these systems by numerically integrating 
the orbits of several hundred test particles, uniformly distributed
along the $x$-axis, in initial circular orbits.
Figure 2 shows the graphs of the lifetimes of these particles for 10 Myr.
As shown here, unlike the stable orbit of the
newly discovered Earth-like planet of GJ 876 
\cite[(Rivera et al., 2005)]{Rivera05}, 
the orbit of the small close-in planet of 55 Cnc, as reported by
\cite[McArthur et al. (2004)]{McArthur04} is unstable.
Our results also indicate that it is unlikely that $\upsilon$ Andromedae 
and GJ 876 harbor habitable planets. This has also been confirmed
by the direct integration of the orbit of an Earth-sized object
in the habitable zone of the system by 
\cite[Dove \& Haghighipour (2006)]{dove06}.
The two systems of 47 UMa and 55 Cnc, however, have stable habitable zones,
although direct integrations of actual Earth-like objects in these systems
are necessary to confirm their habitability.
The results of our test particle simulations also indicated the
capability of 55 Cnc system in harboring stable planet(s) in the
region between 0.7 AU and 2.2 AU. As shown by 
\cite[Fischer et al. (2007)]{Fischer07} 
and as in figure 2, the newly discovered Neptune-sized planet of this system is
located in this region.

\begin{table}
  \begin{center}
\vspace*{1.0 cm}
  \caption{Extrasolar Planet-Hosting Stars in Binary Systems
\cite[(Haghighipour 2006)]{Hagh06}}
  \label{tab1}
 {\scriptsize
  \begin{tabular}{|l|l|l|l|}\hline 
{\bf Star} & {\bf Star} & {\bf Star} & {\bf Star} \\ \hline
HD142    (GJ 9002)                & 
HD3651                            & 
HD9826   ($\upsilon$ And)         &    
HD13445  (GJ 86)                  \\
HD19994                           & 
HD22049  ($\epsilon$ Eri)         & 
HD27442                           & 
HD40979                           \\
HD41004                           & 
HD75732  (55 Cnc)                 & 
HD80606                           & 
HD89744                           \\
HD114762                          & 
HD117176 (70 Vir)                 & 
HD120136 ($\tau$ Boo)             &  
HD121504                          \\
HD137759                          & 
HD143761 ($\rho$ Crb)             & 
HD178911                          & 
HD186472 (16 Cyg)                 \\
HD190360 (GJ 777A)                & 
HD192263                          & 
HD195019                          & 
HD213240                          \\
HD217107                          & 
HD219449                          & 
HD219542                          & 
HD222404 ($\gamma$ Cephei)        \\
HD178911                          &
HD202206                          &
PSR B1257-20                      &
PSR B1620-26                      \\ 
\hline
\end{tabular}}
\end{center}
\end{table}

\begin{figure}
\vspace*{0.7 cm}
\begin{center}
\includegraphics[width=2.6in]{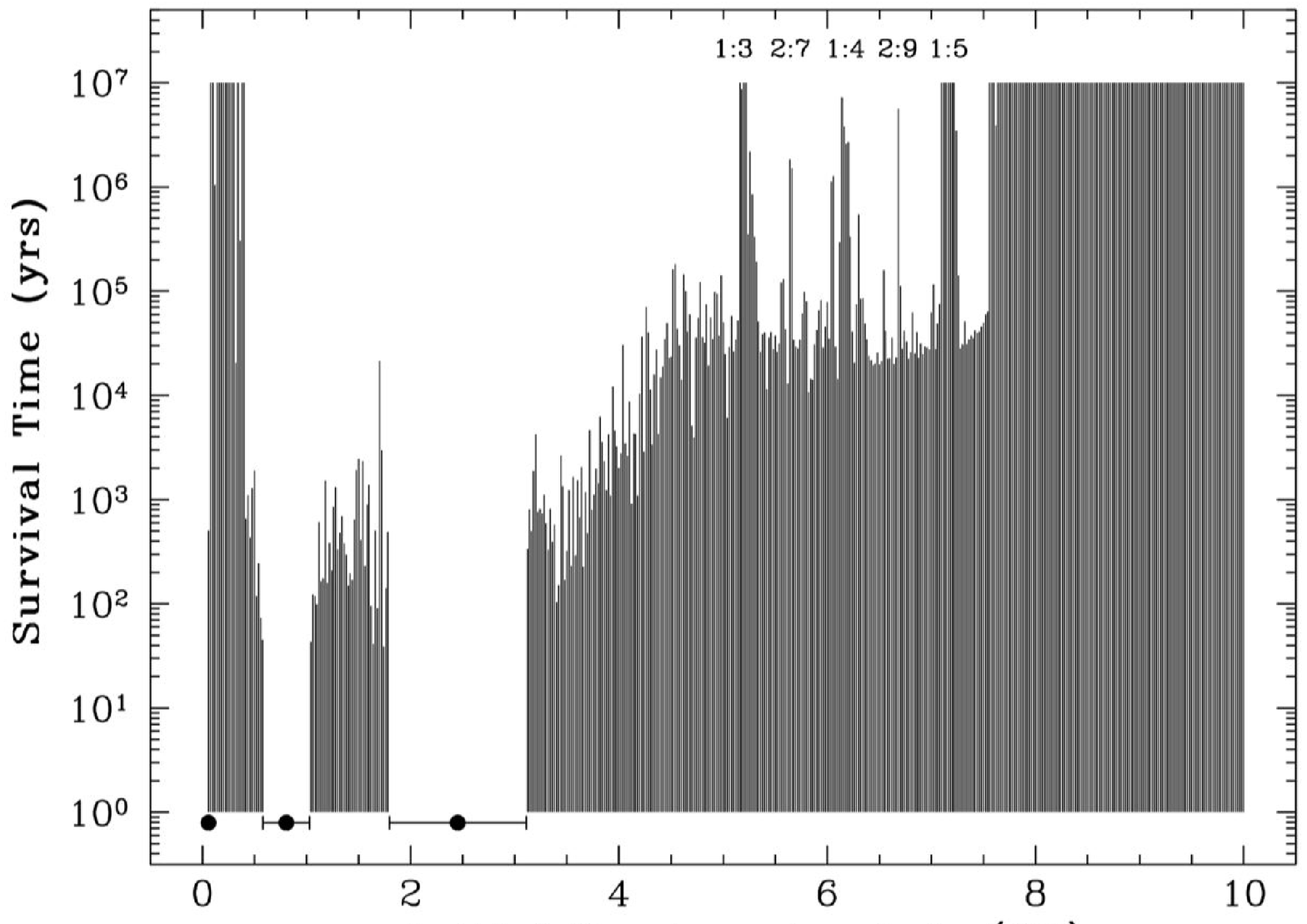}
\includegraphics[width=2.6in]{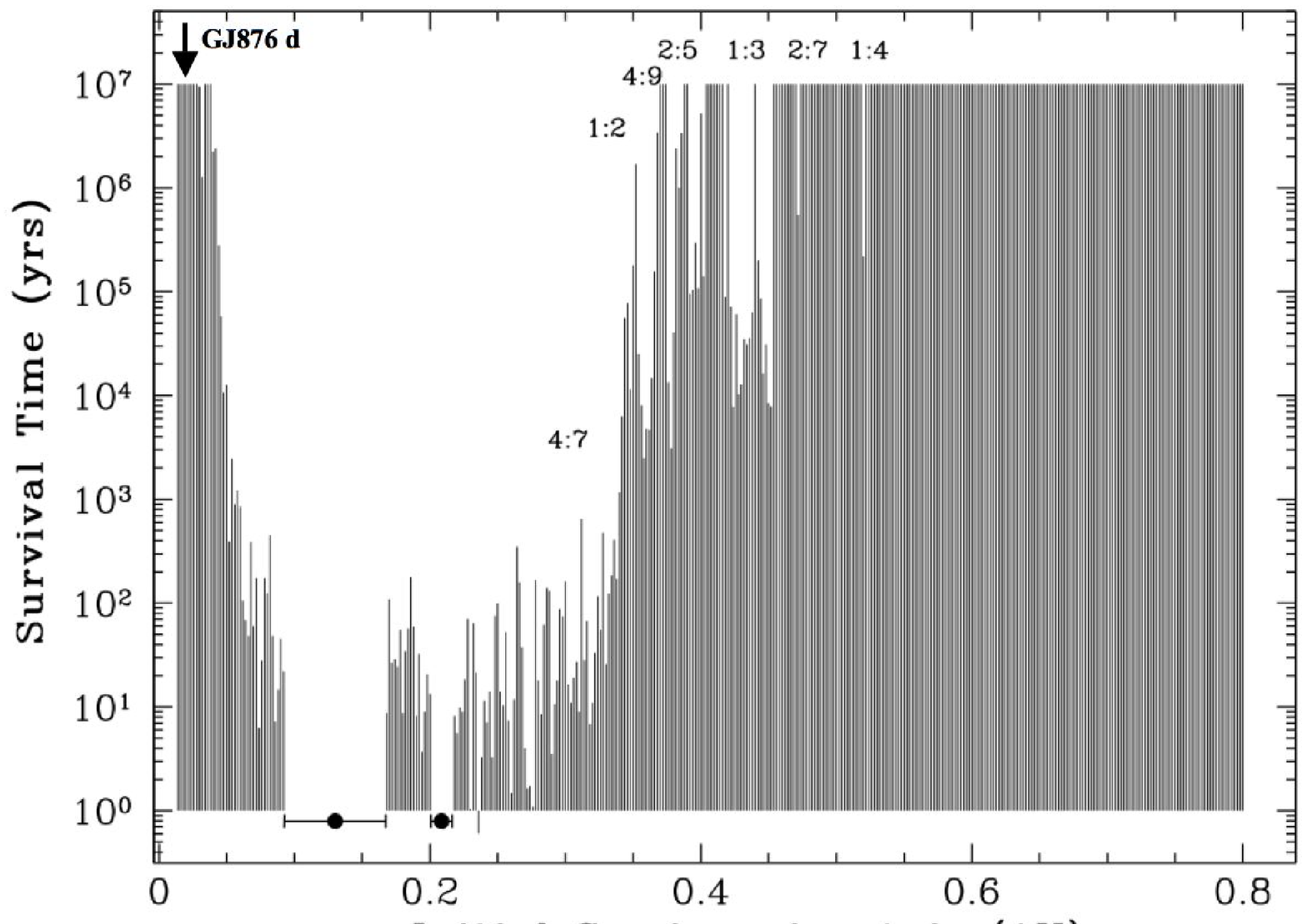}
\includegraphics[width=2.6in]{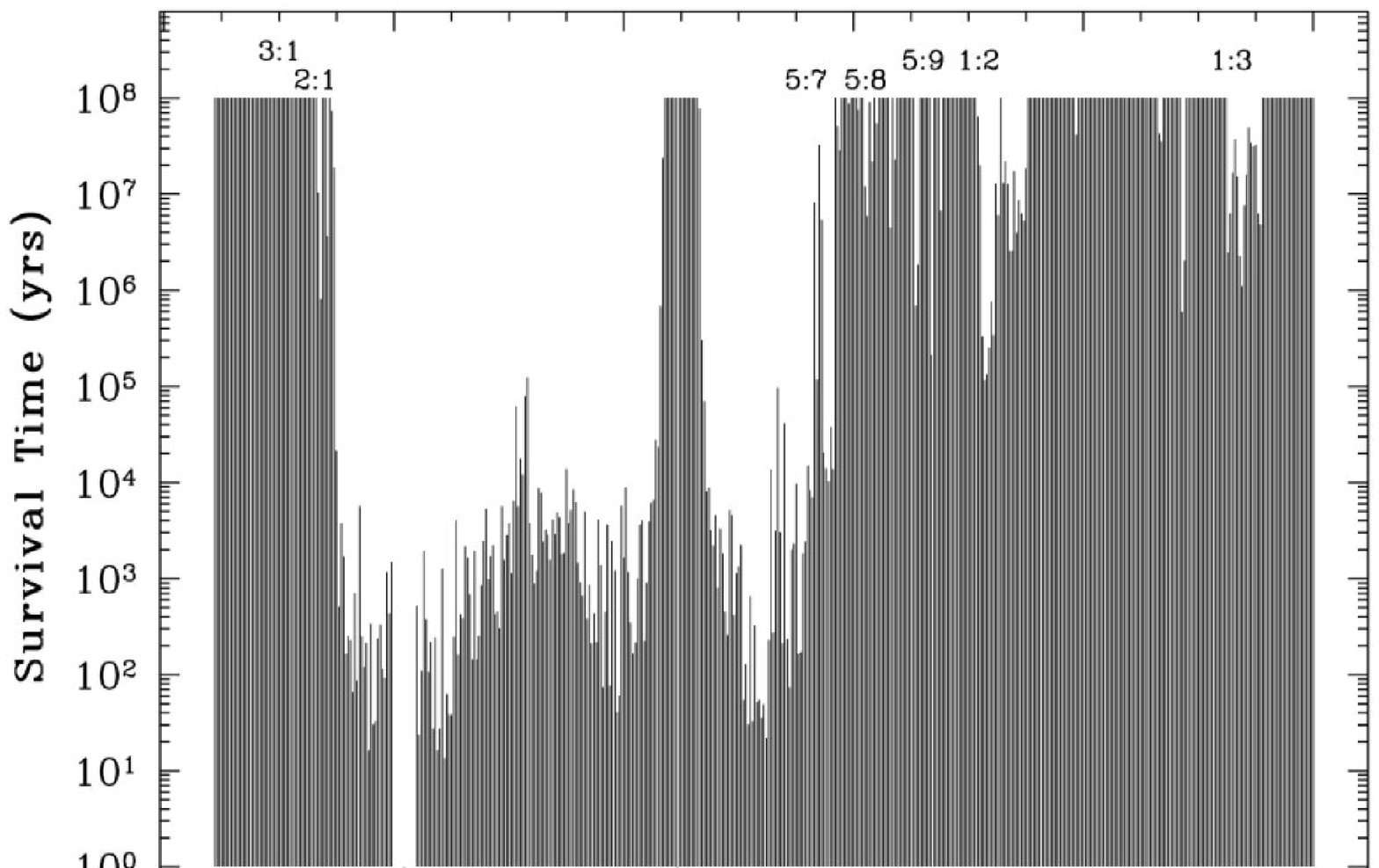}
\includegraphics[width=2.6in]{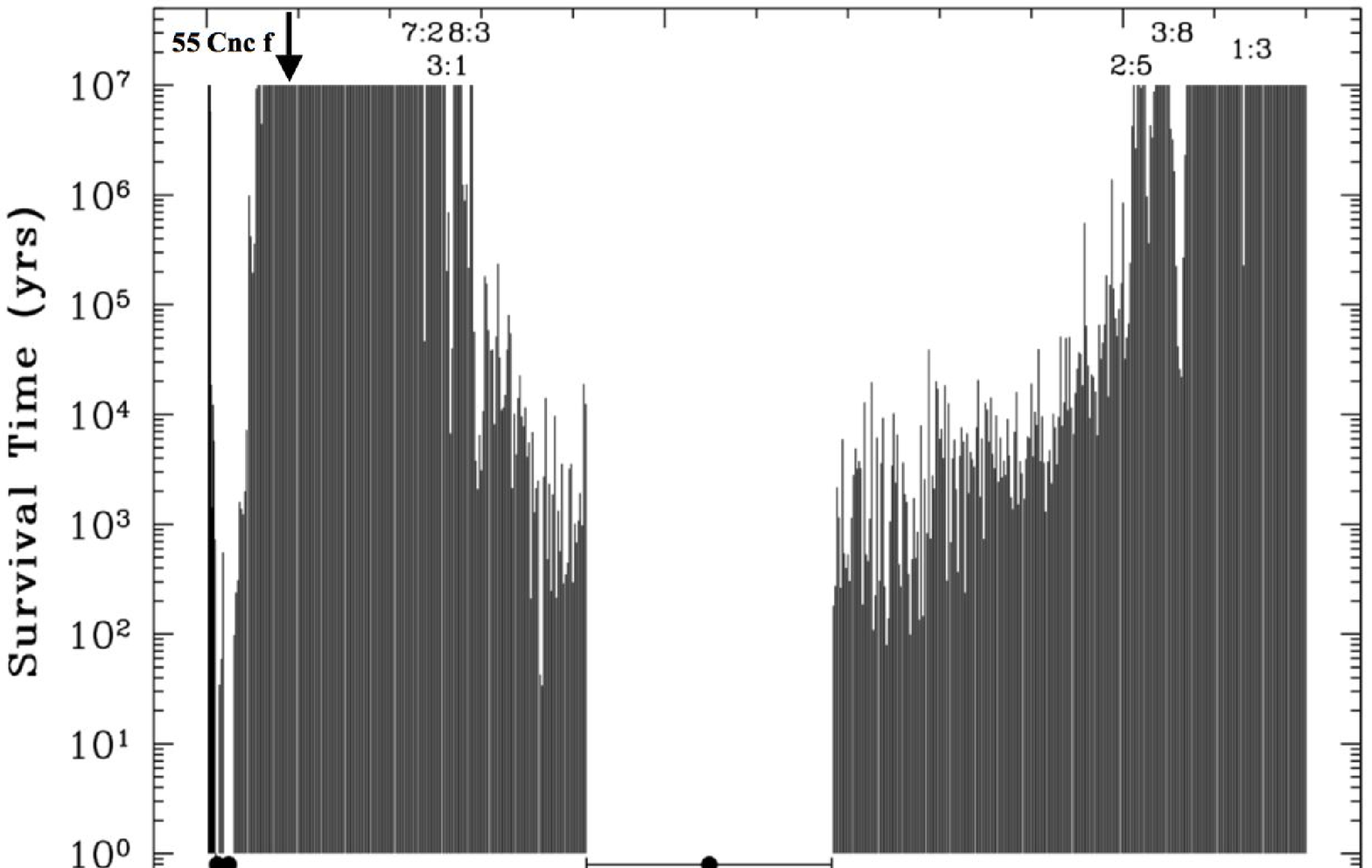}
\end{center}
\vspace*{1.0 cm}
\caption{Graphs of the lifetimes of test particles in $\upsilon$ Andromedae
(top left, HZ=1.68-2 AU), GJ 876 (top right, HZ=0.1-0.3 AU), 
47 UMa (bottom left, HZ=1.16-1.41), and 55 Cnc (bottom right-0.72-0.87 AU).
The graphs and habitable zones are from 
\cite[Rivera \& Haghighipour (2007)]{Rivera07}.
The islands of stability and instability, with their corresponding 
mean-motion resonances
with the inner and/or outer planet are also shown. As shown here, 
the habitable zones of
$\upsilon$ Andromedae and GJ 876 are unstable implying that the 
planetary systems
of these stars will not be habitable. The habitable zones of 
47 UMa and 55 Cnc,
on the other hand, are stable. Also as shown here, 
the recently detected Earth-like planet of GJ 876 
\cite[(Rivera et al. 2005)]{Rivera05}, 
and the newly discovered fifth planet of 55 Cnc
\cite[(Fischer et al. 2007)]{Fischer07} are in stable orbits. 
\label{fig2}}
\end{figure}

\section{Habitability of Multiple Star Systems}

As shown in Table 1, more than 20\% of currently known planet-hosting
stars are members of binary systems \cite[(Haghighipour 2006)]{Hagh06}. 
Many of these systems are wide
with separations ranging from 200 AU to 6000 AU. In such systems, 
the perturbative 
effect of the stellar companion is negligible and planet formation 
around the other
star may proceed in the similar fashion as around a single star. 
There are, however, 
three binary systems, namely, GL 86 \cite[(Els et al 2001)]{Els01}, 
$\gamma$ Cephei \cite[(Hatzes et al 2003)]{Hatzes03}, 
and HD 41004 \cite[(Zucker et al 2004, Raghavan et al 2006)]{Zucker04,Ragh06}, 
in which the primary star is host to a Jovian-type
planet and the binary separation is smaller than 20 AU. How 
these planets were formed, 
and whether such {\it binary-planetary} systems can be habitable 
are now among major
theoretical challenges of planetary dynamics.

Planet formation in close binary systems is strongly affected 
by the perturbation of the
binary companion. This star may remove planet-forming material 
by truncating the primary's  
circumstellar disk \cite[(Artymowicz \& Lubow 1994)]{Artymowicz94} and 
destabilizing the regions where planetesimals and protoplanets
may under go collisional growth \cite[(Th\'ebault et al 2004)]{Thebault04}. 
In binary systems where the primary hosts a giant planet, 
the perturbative effect of the planetary companion will also 
affect the growth of protoplanetary
objects. However, as shown by numerical integrations of the orbits of 
Earth-sized planets in 
$\gamma$ Cephei system  \cite[(Haghighipour 2006)]{Hagh06}, 
it is possible for a terrestrial-class body to maintain
a long-term stable orbit at distances close to the primary star 
and outside the giant planet's
influence zone. Figure 3 shows the graph of the lifetime of 
an Earth-sized object in
the system of $\gamma$ Cephei. 
As shown here, the HZ of the system is unstable. However,
an Earth-like planet can main a stable orbit close  
to the primary star.

\begin{figure}
\vspace*{0.5 cm}
\begin{center}
\includegraphics[width=4in]{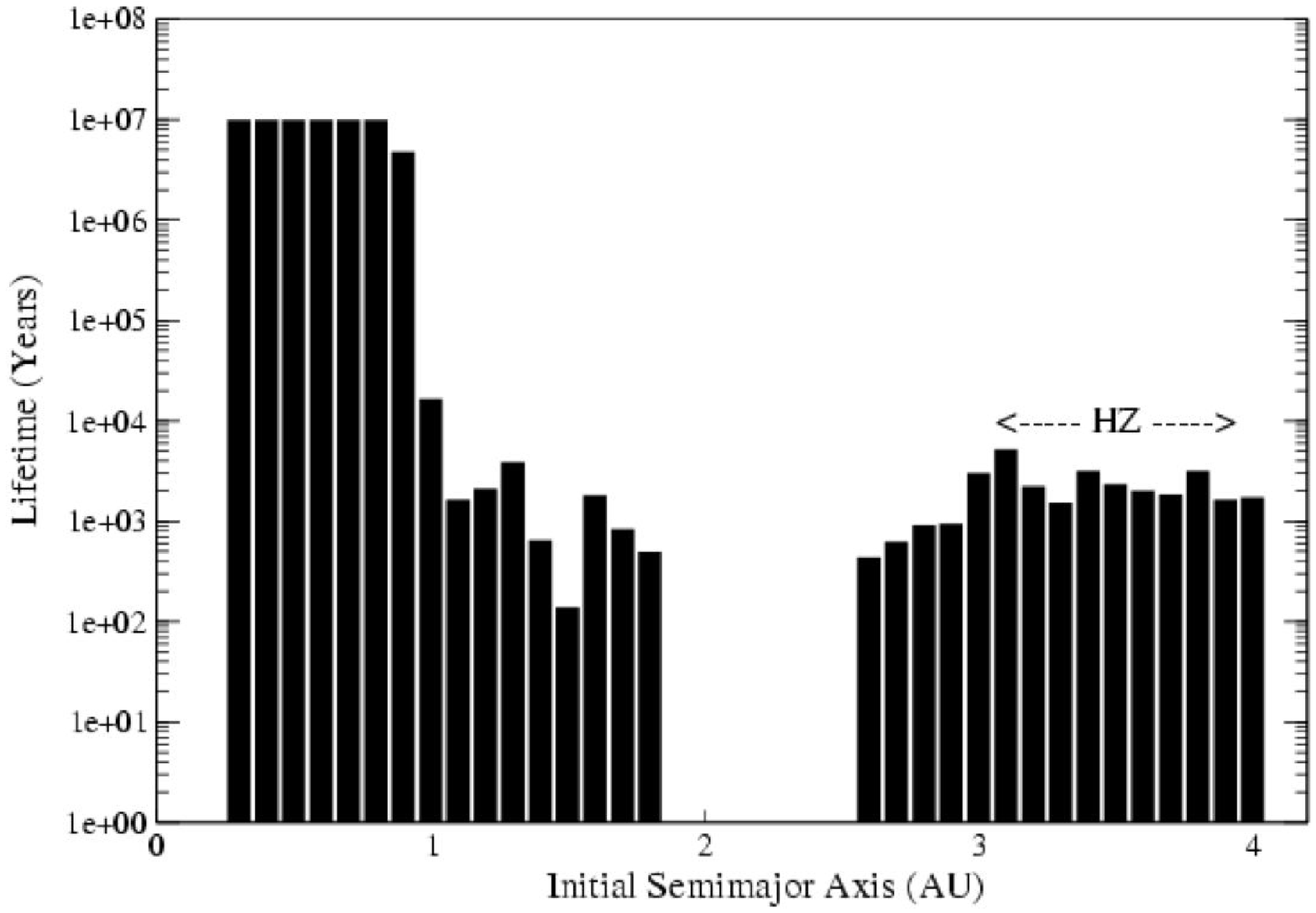} 
\end{center}
\caption{Lifetime of an Earth-size planet in $\gamma$ Cephei system. 
The giant planet of the system (1.67 Jupiter-mass) is at 2.13 AU
with an eccentricity of 0.12. As shown here, the HZ of the system 
is unstable. However,
a terrestrial-class object can maintain a long-term orbit at close distances
to the primary star \cite[(Haghighipour 2006)]{Hagh06}.
\label{fig3}}
\end{figure}

\begin{figure}
\vspace*{0.5 cm}
\begin{center}
\includegraphics[width=4in]{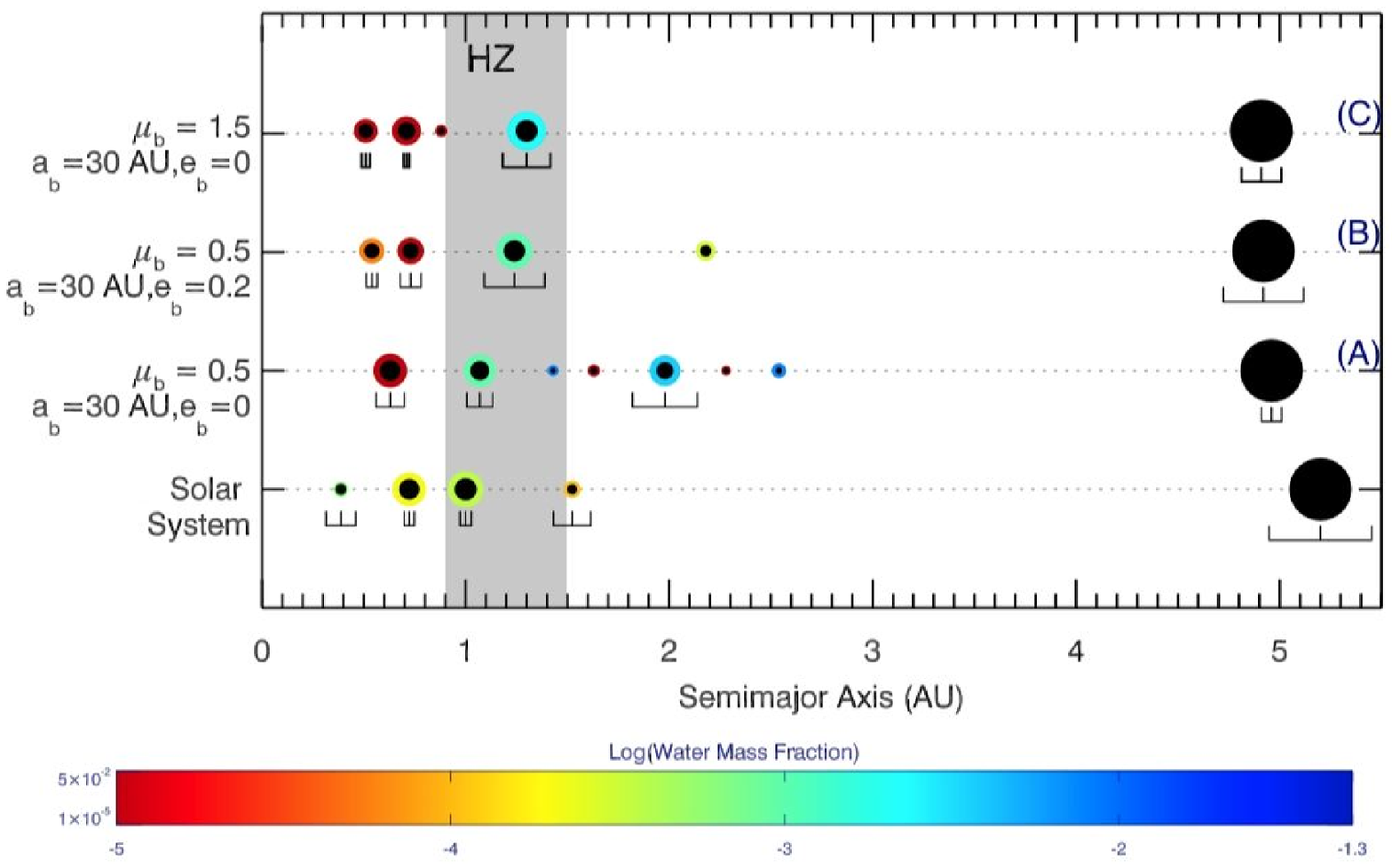} 
\end{center}
\caption{Habitable planet formation in binary-planetary systems 
with 0.5 and 1.5 
stellar mass-ratios. As shown here, binaries with moderate 
periastron distances
are more favorable for the formation of terrestrial-class planets with
considerable amount of water \cite[(Haghighipour \& Raymond 2007)]{Hagh07}.
\label{fig4}}
\end{figure}

Based on the results of the simulations shown in figure  3, we recently
studied habitable planet formation in moderately closed binary star systems
that host giant planets \cite[(Haghighipour \& Raymond 2007)]{Hagh07}. 
We simulated the late stage of terrestrial planet formation
for different values of the semimajor axis and orbital eccentricity of 
the binary,
as well as different binary mass-ratios. Our system consisted of a Sun-like
star as the primary, a disk of protoplanetary bodies with 120 Moon- 
to Mars-sized
objects distributed randomly between 0.5 AU and 4 AU, and a 
Jupiter-sized planet at 5 AU.
To study the effect of the orbital dynamics of the secondary star 
on the formation
of planets in the HZ of the primary and their water contents,  we considered
the orbit of the giant planet to be circular and assumed that the 
distribution of water in 
the protoplanetary disk is similar to those of the primitive asteroids in
the asteroid belt. Figure 4 shows some of the results for binary mass-ratios
${\mu_b}=0.5, 1.5$. As shown here, it is possible
to form Earth-like objects with substantial amount of water 
in the HZ of the primary
star. The sizes of these planets and their water contents
vary with the semimajor axis and eccentricity of the stellar companion. 
In binaries where the secondary star has a small periastron, the interaction
between this object and the giant planet of the system, which transfers 
angular momentum
to the disk of planetary embryos, causes many of these bodies to be ejected
from the system. As a result, in closer and 
eccentric binaries, the final planets are
smaller and contain less or no water. Figure 5 shows the relation between 
the periastron of the binary $(q_b)$ and the
semimajor axis of the outermost terrestrial planet $(a_{out})$. 
As shown in the left graph of figure 5, similar to 
\cite[Quintana et al (2007)]{Quintana07},
simulations with no giant planets favor regions interior to 
$0.19{q_b}$ for the formation of terrestrial objects.
That means, around a Sun-like star, where the inner edge of the
habitable zone is at $\sim 0.9$ AU, 
a stellar companion with a perihelion distance smaller than 
0.9/0.19 = 4.7 AU would not allow habitable planet formation.
In simulations with
giant planets, on the other hand, figure 5 shows that 
terrestrial planets form
closer-in. The ratio ${a_{out}}/{q_b}$ in these systems is between 
0.06 and 0.13.
A detailed analysis of our simulations also indicate
that the systems, in which habitable planets were formed, have
large periastra. The right graph of figure 5 shows this 
for simulations
in a binary with equal-mass Sun-like stars. The circles in this 
figure represent systems with habitable planets.
The numbers on the top of the circles show the mean 
eccentricity of the giant planet. For comparison, systems 
with unstable giant planets have also been marked. Since at the 
beginning of each
simulation, the orbit of the giant planet was considered to be 
circular, a non-zero
eccentricity is indicative of the interaction of this body with 
the secondary star.
As shown here, Earth-like objects are formed in systems where 
the interaction
between the giant planet and the secondary star is weak
and the  average eccentricity of the giant planet is small. 
That implies,
habitable planet formation is more favorable in binaries with 
moderate to large
perihelia, and with giant planets on low eccentricity orbits.

\begin{figure}
\vspace*{0.5 cm}
\begin{center}
\includegraphics[width=2.5in]{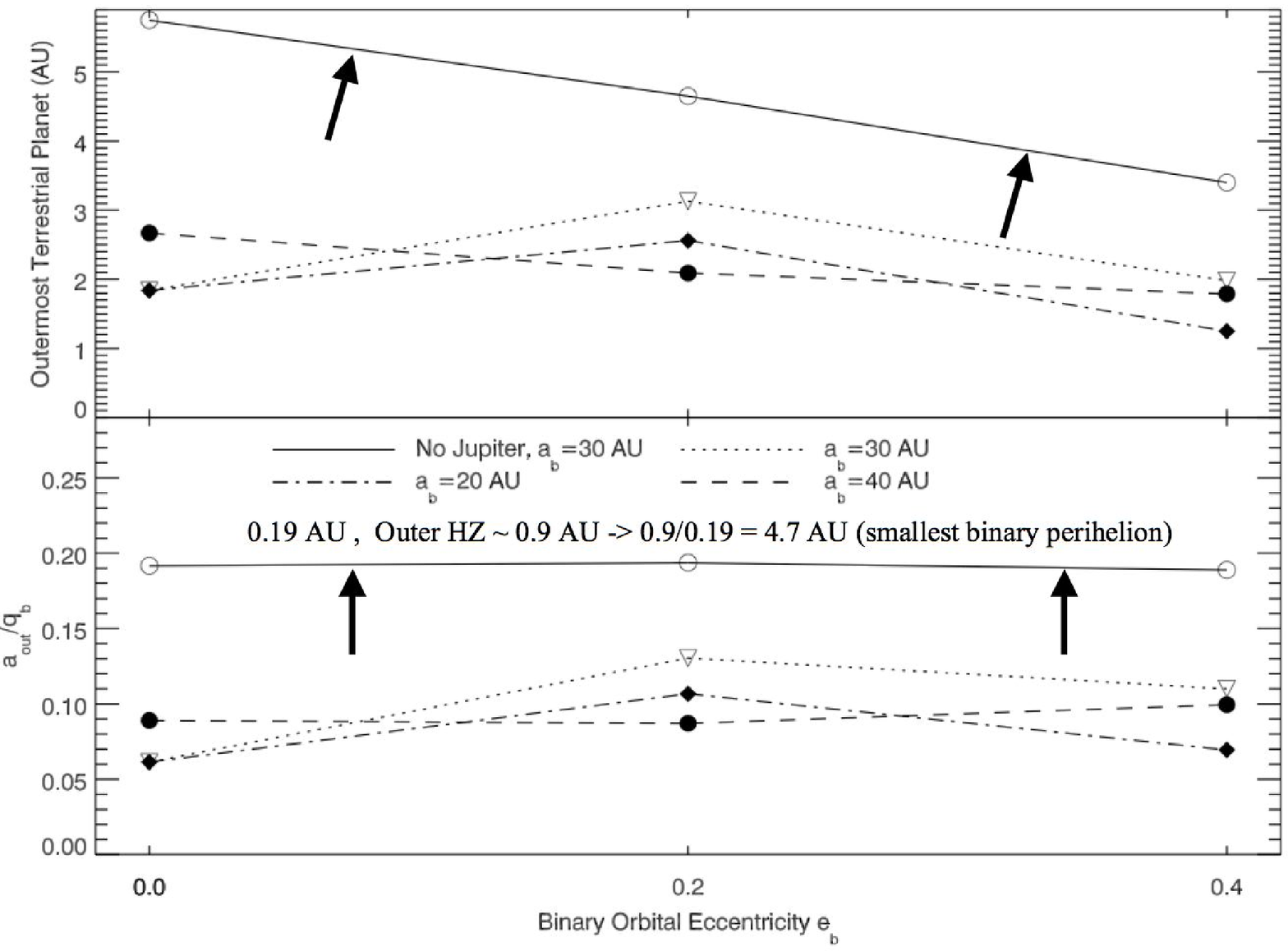} 
\includegraphics[width=2.5in]{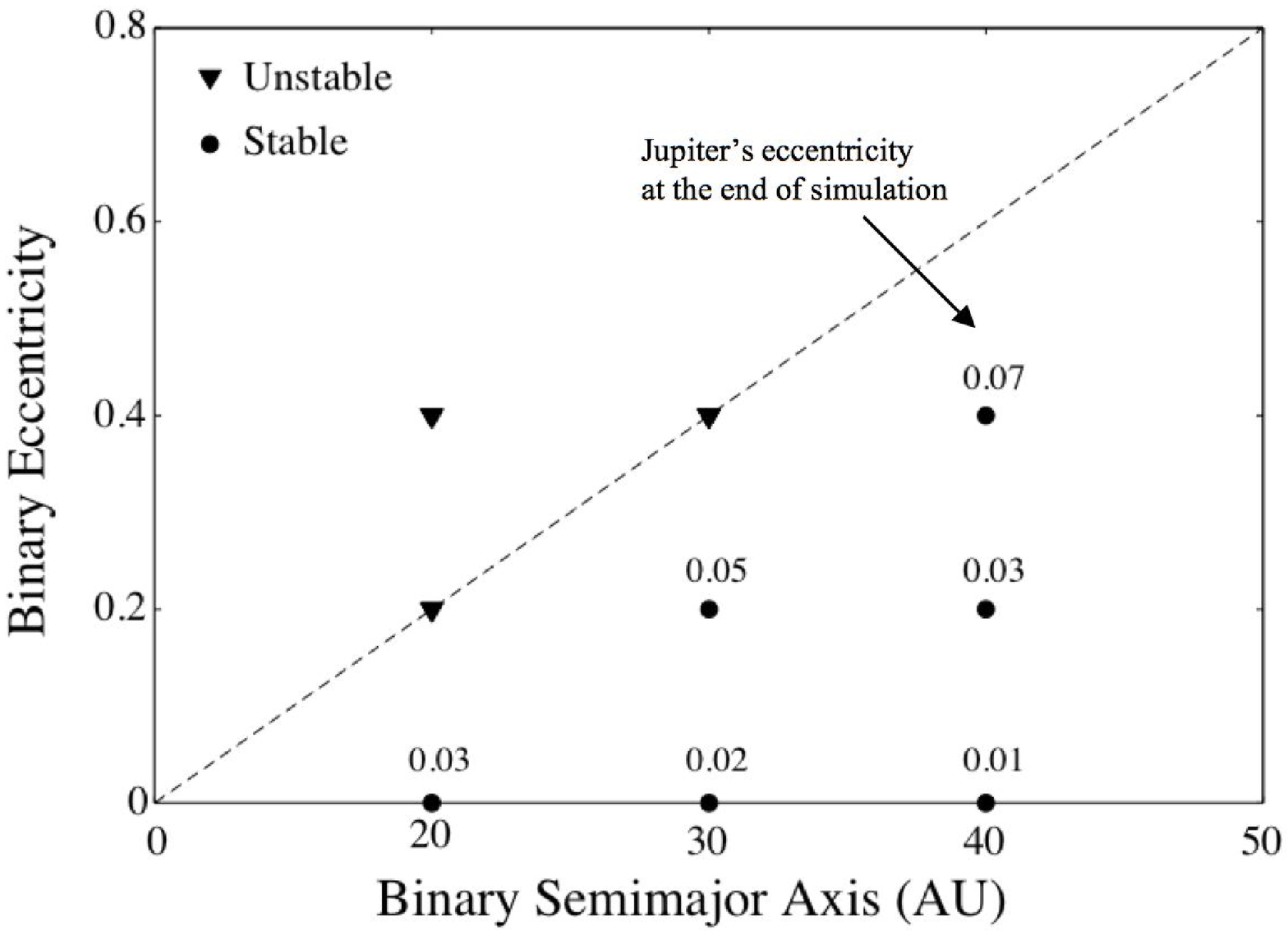}
\end{center}
\caption{The graph on the left shows the relation between the periastron
of an equal-mass binary and the location of its 
outermost terrestrial planet. 
The graph on the right shows the region of the $({e_b},{a_b})$ space
for a habitable binary-planetary system 
\cite[(Haghighipour \& Raymond 2007)]{Hagh07}.
\label{fig5}}
\end{figure}

\acknowledgements
Support by the NASA Astrobiology Institute under Cooperative Agreement 
NNA04CC08A with the Institute for Astronomy at the University of
Hawaii-Manoa is acknowledged.

\end{document}